\documentclass[12pt,a4paper]{article}
\usepackage{amsmath}
\usepackage{latexsym}
\makeatletter
\def\rddots{\mathinner{\mkern1mu\raise\p@%
    \vbox{\kern7\p@\hbox{.}}\mkern2mu%
    \raise4\p@\hbox{.}\mkern2mu\raise7\p@\hbox{.}\mkern1mu}}
\makeatother
\newcommand{\ket}[1]{{\vert{#1}\rangle}}

\newcommand{\fukuso}{{\mathbf C}}

\begin{document}

\title{\sl Introduction to the Rotating Wave Approximation (RWA) : 
Two Coherent Oscillations}
\author{
  Kazuyuki FUJII
  \thanks{E-mail address : fujii@yokohama-cu.ac.jp }\\
  International College of Arts and Sciences\\
  Yokohama City University\\
  Yokohama, 236--0027\\
  Japan
  }
\date{}
\maketitle
\begin{abstract}
  In this note I introduce a mysterious approximation 
called the rotating wave approximation (RWA) to undergraduates 
or non--experts who are interested in both Mathematics and 
Quantum Optics. 

In Quantum Optics it plays a very important role in order to obtain 
an analytic approximate solution of some Schr{\"o}dinger equation, 
while it is curious from the mathematical point of view. 

I explain it carefully with two coherent oscillations for them 
and expect that they will overcome the problem in the near future.
\end{abstract}

\vspace{5mm}\noindent

\vspace{5mm}\noindent

\section{Introduction}
When undergraduates study Quantum Mechanics they encounter 
several approximation methods like the WKB, the Born--Oppenheimer, 
the Hartree--Fock, etc. In fact, exactly solvable models are very few 
in Quantum Mechanics, so (many) approximation methods play 
an important role. As a text book of Quantum Mechanics 
I recommend \cite{HG} although it is not necessarily standard.

When we study Quantum Optics we again encounter the same 
situation. We often use a method called 
the rotating wave approximation (RWA), which means fast 
oscillating terms (in effective Hamiltonians) removed. Because 
\[
e^{\pm i n\theta}\Longrightarrow 
\int e^{\pm i n\theta}d\theta
=\frac{e^{\pm i n\theta}}{\pm in}\approx 0 
\]
holds if $n$ is large enough. We believe that there is no problem on 
this approximation.

However, in some models slow oscillating terms are removed. 
Let us show an example.  The Euler formula gives
\[
e^{i\theta}=\cos{\theta}+i\sin{\theta}\ 
\Longrightarrow\  
2\cos{\theta}=e^{i\theta}+e^{-i\theta}.
\]
From this we approximate $2\cos{\theta}$ to be
\[
2\cos{\theta}
=e^{i\theta}+e^{-i\theta}
=e^{i\theta}(1+e^{-2i\theta})
\approx e^{i\theta}
\]
because $e^{-i\theta}$ goes away from $e^{i\theta}$ 
by two times speed, so we neglect this term. 
In our case $n$ is $2$ ! Read the text for more details.

Why is such a ``rude" method used ? The main reason 
is to obtain {\bf analytic} approximate solutions for some 
important models in Quantum Optics. 
To the best of our knowledge we cannot obtain such 
analytic solutions without RWA.

In this review note I introduce the rotating wave approximation 
in details with two models for undergraduates or non--experts. 
I expect that they will overcome this ``high wall" 
in the near future.

\section{Principles of Quantum Mechanics}
One of targets of the paper is to study and solve the time evolution 
of a quantum state (which is a superposition of two physical states).

In order to set the stage and to introduce proper notation, 
let us start with a system of principles of Quantum 
Mechanics (QM in the following for simplicity). See for example 
\cite{HG}, \cite{PD}, \cite{AP} and \cite{AH}. That is,

\vspace{5mm}\noindent
\begin{Large}
{\bf System of Principles of QM}
\end{Large}

\vspace{3mm}\noindent
1.\ {\bf Superposition Principle}\\
If $\ket{a}$ and $\ket{b}$ are physical states then their 
superposition $\alpha\ket{a}+\beta\ket{b}$ is also 
a physical state where $\alpha$ and $\beta$ are 
complex numbers.

\vspace{3mm}\noindent
2.\ {\bf Schr\"{o}dinger Equation and Evolution}\\
Time evolution of a physical state proceeds like
\[
\ket{\Psi}\ \longrightarrow\ U(t)\ket{\Psi}
\]
where $U(t)$ is the unitary evolution operator 
($U^{\dagger}(t)U(t)=U(t)U^{\dagger}(t)={\bf 1}$ and $U(0)={\bf 1}$) 
determined by a Schr\"{o}dinger Equation.

\vspace{3mm}\noindent
3.\ {\bf Copenhagen Interpretation}\footnote{There are some 
researchers who are against this terminology, see for 
example \cite{AH}. However, I don't agree with them 
because the terminology is nowadays very popular in the 
world}\\
Let $a$ and $b$ be the eigenvalues of an observable $Q$, and 
$\ket{a}$ and $\ket{b}$ be the normalized eigenstates corresponding to 
$a$ and $b$. When a state is a superposition $\alpha\ket{a}+\beta\ket{b}$ 
and we observe the observable $Q$ the state collapses like
\[
\alpha\ket{a}+\beta\ket{b}\ \longrightarrow \ \ket{a}
\quad\mbox{or}\quad
\alpha\ket{a}+\beta\ket{b}\ \longrightarrow \ \ket{b}
\]
where their collapsing probabilities are $|\alpha|^{2}$ and 
$|\beta|^{2}$ respectively ($|\alpha|^{2}+|\beta|^{2}=1$).

This is called the collapse of the wave function and 
the probabilistic interpretation.

\vspace{3mm}\noindent
4.\ {\bf Many Particle State and Tensor Product}\\
A multiparticle state can be constructed by the superposition of 
the Knonecker products of one particle states, which are called 
the tensor products. For example, 
\[
\alpha|{a}\rangle\otimes|{a}\rangle +\beta|{b}\rangle\otimes|{b}\rangle
\equiv 
\alpha|{a,a}\rangle +\beta|{b,b}\rangle
\]
is a two particle state.

\vspace{5mm}
These will play an essential role in the later sections.

\section{Two--Level System of an Atom}
In order to treat the two--level system of an atom 
we make a short review of the two--dimensional complex 
vector space $\fukuso^{2}$ and complex matrix space 
$M(2;\fukuso)$ within our necessity.  
See for example \cite{Five}.
 
First we introduce the (famous) Pauli matrices  
$\{\sigma_{1},\sigma_{2},\sigma_{3}\}$ defined by
\begin{equation}
\label{eq:Pauli}
\sigma_{1}=
\left(
\begin{array}{cc}
0 & 1 \\
1 & 0
\end{array}
\right),\quad
\sigma_{2}=
\left(
\begin{array}{cc}
0 & -i \\
i  & 0
\end{array}
\right),\quad
\sigma_{3}=
\left(
\begin{array}{cc}
1 & 0  \\
0 & -1
\end{array}
\right)
\end{equation}
and set the unit matrx $1_{2}$ by
\[
1_{2}=
\left(
\begin{array}{cc}
1 & 0 \\
0 & 1
\end{array}
\right).
\]
Moreover, we set
\[
\sigma_{+}=
\frac{1}{2}(\sigma_{1}+i\sigma_{2})=
\left(
\begin{array}{cc}
0 & 1 \\
0 & 0
\end{array}
\right),
\quad
\sigma_{-}=
\frac{1}{2}(\sigma_{1}-i\sigma_{2})=
\left(
\begin{array}{cc}
0 & 0 \\
1 & 0
\end{array}
\right).
\]
Note that $\sigma_{1}=\sigma_{+}+\sigma_{-}$. 
Then it is easy to see
\begin{equation}
\label{eq:su(2)-relations}
[\frac{1}{2}\sigma_{3},\sigma_{+}]=\sigma_{+},\quad
[\frac{1}{2}\sigma_{3},\sigma_{-}]=-\sigma_{-},\quad
[\sigma_{+},\sigma_{-}]=2\times \frac{1}{2}\sigma_{3}.
\end{equation}

\vspace{5mm}\noindent
{\bf Comment}\ \ The Pauli matrices 
$\{\sigma_{1},\sigma_{2},\sigma_{3}\}$ are generators of the Lie algebra 
$su(2)$ of the special unitary group $SU(2)$ and 
$\{\sigma_{+},\sigma_{-},\frac{1}{2}\sigma_{3}\}$ are generators of 
the Lie algebra $sl(2;\fukuso)$ of the special linear group 
$SL(2;{\bf C})$. For the sake of readers we write a Lie--diagram 
of these algebras and groups.

\vspace{5mm}
\begin{center}
\unitlength 0.1in
\begin{picture}( 24.1500,  8.9500)(  4.0000,-16.0500)
\put(10.7500,-8.1500){\makebox(0,0){$sl(2;\fukuso)$}}%
\put(29.1500,-7.9500){\makebox(0,0){$SL(2;\fukuso)$}}%
\put(11.6000,-16.0000){\makebox(0,0){$su(2)$}}%
\put(28.7000,-15.9000){\makebox(0,0){$SU(2)$}}%
%
\special{pn 8}%
\special{pa 1606 796}%
\special{pa 2396 796}%
\special{fp}%
\special{sh 1}%
\special{pa 2396 796}%
\special{pa 2328 776}%
\special{pa 2342 796}%
\special{pa 2328 816}%
\special{pa 2396 796}%
\special{fp}%
%
\special{pn 8}%
\special{pa 1620 1600}%
\special{pa 2410 1600}%
\special{fp}%
\special{sh 1}%
\special{pa 2410 1600}%
\special{pa 2344 1580}%
\special{pa 2358 1600}%
\special{pa 2344 1620}%
\special{pa 2410 1600}%
\special{fp}%
%
\special{pn 8}%
\special{pa 1120 1400}%
\special{pa 1120 990}%
\special{fp}%
\special{sh 1}%
\special{pa 1120 990}%
\special{pa 1100 1058}%
\special{pa 1120 1044}%
\special{pa 1140 1058}%
\special{pa 1120 990}%
\special{fp}%
%
\special{pn 8}%
\special{pa 2810 1400}%
\special{pa 2810 990}%
\special{fp}%
\special{sh 1}%
\special{pa 2810 990}%
\special{pa 2790 1058}%
\special{pa 2810 1044}%
\special{pa 2830 1058}%
\special{pa 2810 990}%
\special{fp}%
\end{picture}%

\end{center}

\vspace{5mm}
Next, we define $\{\ket{0},\ket{1}\}$ a basis of $\fukuso^{2}$ 
by use of the Dirac's notation
\begin{equation}
\label{eq:basis}
\ket{0}=
\left(
\begin{array}{c}
1 \\
0
\end{array}
\right),
\quad
\ket{1}=
\left(
\begin{array}{c}
0 \\
1
\end{array}
\right).
\end{equation}
Then, since $\sigma_{1}$ satisfies the relation
\[
\sigma_{1}\ket{0}=\ket{1},\quad \sigma_{1}\ket{1}=\ket{0}
\]
it is called the flip operation.

\vspace{5mm}\noindent
{\bf Note}\ \ If we define $\{\sigma_{+},\sigma_{-},\frac{1}{2}\sigma_{3}\}$ 
as above then $\{\ket{0}, \ket{1}\}$ should be chosen as
\[
\ket{0}=
\left(
\begin{array}{c}
0 \\
1
\end{array}
\right), 
\quad
\ket{1}=
\left(
\begin{array}{c}
1 \\
0
\end{array}
\right)
\]
instead of (\ref{eq:basis}). Because,
\[
\sigma_{-}\ket{0}=0, \quad \sigma_{-}\ket{1}=\ket{0},\quad 
\sigma_{+}\ket{0}=\ket{1}.
\]
However, I use the conventional notations in this note.

\vspace{5mm}
For the later convenience we calculate the exponential map. 
For a square matrix $A$ the exponential map is defined by
\[
e^{\lambda A}
=\sum_{n=0}^{\infty}\frac{(\lambda A)^{n}}{n!}
=\sum_{n=0}^{\infty}\frac{\lambda^{n}}{n!}A^{n}, 
\quad A^{0}=E,
\]
where $E$ is the unit matrix and $\lambda$ is a constant.

Here, let us calculate $e^{i\lambda \sigma_{1}}$ as an 
example. Noting
\[
\sigma_{1}^{2}
=
\left(
\begin{array}{cc}
0 & 1 \\
1 & 0
\end{array}
\right)^{2}
=
\left(
\begin{array}{cc}
1 & 0 \\
0 & 1
\end{array}
\right)
=1_{2}
\]
we obtain
\begin{eqnarray}
\label{eq:exponential}
e^{i\lambda \sigma_{1}}
&=&
\sum_{n=0}^{\infty}\frac{(i\lambda)^{n}{\sigma_{1}}^{n}}{n!}
\nonumber \\
&=&
\sum_{n=0}^{\infty}\frac{(i\lambda)^{2n}}{(2n)!}{\sigma_{1}}^{2n}+
\sum_{n=0}^{\infty}\frac{(i\lambda)^{2n+1}}{(2n+1)!}{\sigma_{1}}^{2n+1}
\nonumber \\
&=&
\sum_{n=0}^{\infty}\frac{(-1)^{n}\lambda^{2n}}{(2n)!}1_{2}+
i\sum_{n=0}^{\infty}\frac{(-1)^{n}\lambda^{2n+1}}{(2n+1)!}\sigma_{1}
\nonumber \\
&=&
\cos \lambda\ 1_{2}+i\sin \lambda\ \sigma_{1}
\nonumber \\
&=&
\left(
\begin{array}{cc}
\cos \lambda & i\sin \lambda \\
i\sin \lambda  & \cos \lambda
\end{array}
\right).
\end{eqnarray}

\vspace{5mm}\noindent
{\bf Exercise}\ \ Calculate
\[
e^{i\lambda \sigma_{2}}\quad \mbox{and}\quad
e^{i\lambda \sigma_{3}}.
\]

We discuss an atom trapped in a cavity and consider only 
two energy states, namely (in our case) the ground state and 
first excited state. That is, all the remaining states are neglected. 
This is usually called the two--level approximation. 
See for example \cite{WS} as a general introduction.

We set that energies of the ground state $\ket{0}$ and 
first excited state $\ket{1}$ are $E_{0}$ and 
$E_{1}$i$E_{0}<E_{1}$jrespectively. 
Under this approximation the space of all states is two--dimensional, 
so there is no problem to identify $\{\ket{0}, \ket{1}\}$ with (\ref{eq:basis}).

Then we can write the Hamiltonian in a diagonal form like
\[
H_{0}=
\left(
\begin{array}{cc}
E_{0} &        \\
       & E_{1}
\end{array}
\right).
\]
For the later convenience let us transform it. 
For $\Delta=E_{1}-E_{0}$ the energy difference we have
\begin{eqnarray}
\left(
\begin{array}{cc}
E_{0} &        \\
       & E_{1}
\end{array}
\right)
&=&
\left(
\begin{array}{cc}
\frac{E_{0}+E_{1}}{2}-\frac{E_{1}-E_{0}}{2} &     \\
  & \frac{E_{0}+E_{1}}{2}+\frac{E_{1}-E_{0}}{2}
\end{array}
\right)   \nonumber \\
&=&
\frac{E_{0}+E_{1}}{2}1_{2}-\frac{E_{1}-E_{0}}{2}\sigma_{3} 
\nonumber \\
&=&\frac{E_{0}+E_{1}}{2}1_{2}-\frac{\Delta}{2}\sigma_{3}.
\end{eqnarray}

To this atom we subject 
LASER ({\bf L}ight {\bf A}mplification 
by {\bf S}timulated {\bf E}mission of {\bf R}adiation) 
in order to controll it. 
As an image see the following figure.

\vspace{3mm}
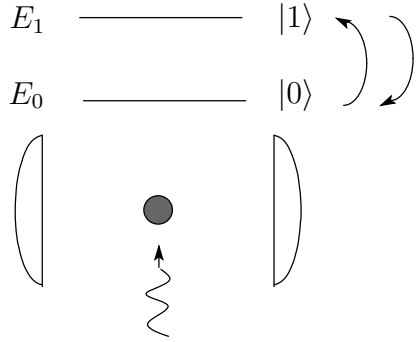
\begin{center}
\unitlength 0.1in
\begin{picture}( 23.1600, 17.4000)(  6.0000,-20.6000)
%
\special{pn 8}%
\special{sh 0.600}%
\special{ar 1596 1400 74 74  0.0000000 6.2831853}%
%
\special{pn 8}%
\special{pa 1210 830}%
\special{pa 2050 830}%
\special{fp}%
\put(9.1500,-8.0000){\makebox(0,0){$E_{0}$}}%
\put(9.2000,-4.0500){\makebox(0,0){$E_{1}$}}%
\put(23.1000,-8.0000){\makebox(0,0){$|0\rangle$}}%
\put(23.1000,-4.0500){\makebox(0,0){$|1\rangle$}}%
%
\special{pn 8}%
\special{pa 2196 1800}%
\special{pa 2196 1010}%
\special{fp}%
%
\special{pn 8}%
\special{pa 2196 1010}%
\special{pa 2226 1020}%
\special{pa 2250 1040}%
\special{pa 2268 1066}%
\special{pa 2282 1096}%
\special{pa 2294 1126}%
\special{pa 2304 1156}%
\special{pa 2312 1186}%
\special{pa 2320 1218}%
\special{pa 2324 1250}%
\special{pa 2328 1282}%
\special{pa 2332 1312}%
\special{pa 2334 1344}%
\special{pa 2336 1376}%
\special{pa 2336 1408}%
\special{pa 2336 1440}%
\special{pa 2332 1472}%
\special{pa 2330 1504}%
\special{pa 2326 1536}%
\special{pa 2322 1568}%
\special{pa 2316 1600}%
\special{pa 2308 1630}%
\special{pa 2300 1662}%
\special{pa 2288 1692}%
\special{pa 2276 1720}%
\special{pa 2260 1748}%
\special{pa 2238 1772}%
\special{pa 2210 1788}%
\special{pa 2196 1790}%
\special{sp}%
%
\special{pn 8}%
\special{pa 996 1800}%
\special{pa 996 1010}%
\special{fp}%
%
\special{pn 8}%
\special{pa 996 1010}%
\special{pa 966 1020}%
\special{pa 942 1040}%
\special{pa 922 1066}%
\special{pa 908 1096}%
\special{pa 896 1126}%
\special{pa 888 1156}%
\special{pa 878 1186}%
\special{pa 872 1218}%
\special{pa 866 1250}%
\special{pa 862 1282}%
\special{pa 860 1312}%
\special{pa 858 1344}%
\special{pa 856 1376}%
\special{pa 856 1408}%
\special{pa 856 1440}%
\special{pa 858 1472}%
\special{pa 860 1504}%
\special{pa 864 1536}%
\special{pa 868 1568}%
\special{pa 876 1600}%
\special{pa 882 1630}%
\special{pa 892 1662}%
\special{pa 902 1692}%
\special{pa 916 1720}%
\special{pa 932 1748}%
\special{pa 954 1772}%
\special{pa 980 1788}%
\special{pa 996 1790}%
\special{sp}%
%
\special{pn 8}%
\special{pa 1190 396}%
\special{pa 2030 396}%
\special{fp}%
%
\special{pn 8}%
\special{pa 1610 1710}%
\special{pa 1642 1734}%
\special{pa 1660 1756}%
\special{pa 1650 1776}%
\special{pa 1616 1794}%
\special{pa 1576 1812}%
\special{pa 1542 1826}%
\special{pa 1530 1840}%
\special{pa 1548 1854}%
\special{pa 1586 1868}%
\special{pa 1624 1882}%
\special{pa 1650 1898}%
\special{pa 1644 1916}%
\special{pa 1618 1936}%
\special{pa 1582 1958}%
\special{pa 1552 1978}%
\special{pa 1538 1998}%
\special{pa 1550 2016}%
\special{pa 1580 2034}%
\special{pa 1620 2050}%
\special{pa 1650 2060}%
\special{sp}%
%
\special{pn 8}%
\special{pa 1600 1700}%
\special{pa 1600 1590}%
\special{fp}%
\special{sh 1}%
\special{pa 1600 1590}%
\special{pa 1580 1658}%
\special{pa 1600 1644}%
\special{pa 1620 1658}%
\special{pa 1600 1590}%
\special{fp}%
%
\special{pn 8}%
\special{ar 2556 634 120 220  4.7123890 6.2831853}%
\special{ar 2556 634 120 220  0.0000000 1.5707963}%
%
\special{pn 8}%
\special{pa 2546 414}%
\special{pa 2520 400}%
\special{fp}%
\special{sh 1}%
\special{pa 2520 400}%
\special{pa 2572 448}%
\special{pa 2568 424}%
\special{pa 2590 412}%
\special{pa 2520 400}%
\special{fp}%
%
\special{pn 8}%
\special{ar 2796 610 120 220  4.7123890 6.2831853}%
\special{ar 2796 610 120 220  0.0000000 1.5707963}%
%
\special{pn 8}%
\special{pa 2776 840}%
\special{pa 2750 854}%
\special{fp}%
\special{sh 1}%
\special{pa 2750 854}%
\special{pa 2820 842}%
\special{pa 2798 830}%
\special{pa 2802 806}%
\special{pa 2750 854}%
\special{fp}%
\end{picture}%

\end{center}

\noindent
In this note we treat Laser as a classical wave for simplicity, 
which is not so bad as shown in the following. That is, we may 
set the laser field as
\[
A\cos(\omega t+\phi).
\]

By the way, from several experiments we know that an atom 
subjected by Laser raises an energy level and vice versa. 
This is expressed by the property of the Pauli matrix 
$\sigma_{1}$
\[
\sigma_{1}\ket{0}=\ket{1},\quad \sigma_{1}\ket{1}=\ket{0},
\]
so we can use $\sigma_{1}$ as the interaction term of 
the Hamiltonian.

As a result our Hamiltonian (effective Hamiltonian) can be 
written as
\begin{equation}
\label{eq:Hamiltonian}
H
=-\frac{\Delta}{2}\sigma_{3}+2g\cos(\omega t+\phi)\sigma_{1}
=
\left(
\begin{array}{cc}
-\frac{\Delta}{2}          & 2g\cos(\omega t+\phi)  \\
2g\cos(\omega t+\phi) & \frac{\Delta}{2}
\end{array}
\right)
\end{equation}
where $g$ is a coupling constant regarding an interaction of 
between an atom and laser, and $A$ is absorbed in $g$ 
($gA \longrightarrow g$). We ignore the scalar term 
$\frac{E_{0}+E_{1}}{2}1_{2}$ for simplicity. 
Note that (\ref {eq:Hamiltonian}) is semi--classical and 
time--dependent. 

Therefore, our task is to solve the Schr{\"o}dinger equation
\begin{equation}
\label{eq:Schrodinger}
i\hbar \frac{\partial}{\partial t}\Psi = H\Psi
\end{equation}
exactly (if possible).

\section{Rotating Wave Approximation}
Unfortunately we cannot solve (\ref{eq:Schrodinger}) exactly 
at the present time. It must be non--integrable although we 
don't know the proof (see the appendix). Therefore we must 
apply some approximate method in order to obtain an {\bf analytic} 
approximate solution. Now we explain a method called the 
Rotating Wave Approximation (RWA). Let us recall the 
Euler formula
\[
e^{i\theta}=\cos{\theta}+i\sin{\theta}\ 
\Longrightarrow\  
2\cos{\theta}=e^{i\theta}+e^{-i\theta}.
\]
From this we approximate $2\cos{\theta}$ to be
\begin{equation}
\label{eq:a kind of RWA}
2\cos{\theta}
=e^{i\theta}+e^{-i\theta}
=e^{i\theta}(1+e^{-2i\theta})
\approx e^{i\theta}
\end{equation}
because $e^{-i\theta}$ goes away from $e^{i\theta}$ 
by two times speed, so we neglect this term ! 
We call this the rotating wave approximation.

\vspace{5mm}
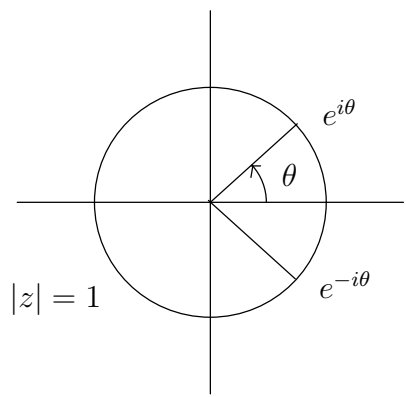
\begin{center}
\unitlength 0.1in
\begin{picture}( 20.4000, 20.0000)( 11.6000,-22.1000)
%
\special{pn 8}%
\special{pa 1200 1210}%
\special{pa 3200 1210}%
\special{fp}%
%
\special{pn 8}%
\special{pa 2200 210}%
\special{pa 2200 2210}%
\special{fp}%
%
\special{pn 8}%
\special{ar 2200 1210 600 600  0.0000000 6.2831853}%
%
\special{pn 8}%
\special{pa 2650 800}%
\special{pa 2200 1210}%
\special{fp}%
%
\special{pn 8}%
\special{pa 2640 1610}%
\special{pa 2190 1200}%
\special{fp}%
%
\special{pn 8}%
\special{ar 2200 1210 290 290  5.5477455 6.2831853}%
\put(26.1000,-10.7000){\makebox(0,0){$\theta$}}%
%
\special{pn 8}%
\special{pa 2420 1020}%
\special{pa 2410 1080}%
\special{fp}%
%
\special{pn 8}%
\special{pa 2410 1020}%
\special{pa 2460 1020}%
\special{fp}%
\put(11.6000,-18.0000){\makebox(0,0)[lb]{$|z|=1$}}%
\put(28.7000,-7.4000){\makebox(0,0){$e^{i\theta}$}}%
\put(29.0000,-16.4000){\makebox(0,0){$e^{-i\theta}$}}%
\end{picture}%

\end{center}

\vspace{3mm}\noindent
{\bf Problem}\ \ In general, fast oscillating terms may 
be neglected because
\[
\int e^{\pm i n\theta}d\theta
=\frac{e^{\pm i n\theta}}{\pm in}\approx 0
\]
if $n$ is large. Our question is : Is $n=2$ large enough ?

\vspace{5mm}
By noting that the Hamiltonian should be hermitian, 
we approximate
\[
2\cos(\omega t+\phi)\sigma_{1}
=
\left(
\begin{array}{cc}
0                             & 2\cos(\omega t+\phi) \\
2\cos(\omega t+\phi) & 0
\end{array}
\right)
\approx 
\left(
\begin{array}{cc}
0                            & e^{i(\omega t+\phi)} \\
e^{-i(\omega t+\phi)} & 0
\end{array}
\right),
\]
by use of (\ref{eq:a kind of RWA}), so (\ref{eq:Hamiltonian}) 
is reduced to
\begin{equation}
\label{eq:reduced Hamiltonian}
\widetilde{H}=
\left(
\begin{array}{cc}
-\frac{\Delta}{2}          & ge^{i(\omega t+\phi)}  \\
ge^{-i(\omega t+\phi)}  & \frac{\Delta}{2} 
\end{array}
\right).
\end{equation}

As a result our modified task is to solve the Schr{\"o}dinger 
equation
\begin{equation}
\label{eq:reduced Schrodinger}
i\hbar \frac{\partial}{\partial t}\Psi = \widetilde{H}\Psi
\end{equation}
exactly. Mysteriously enough, this equation can be solved easily.

\vspace{3mm}\noindent
{\bf Note}\ \ For the latter convenience let us rewrite 
the method with formal notations :
\begin{eqnarray*}
2\cos{\theta}\ \sigma_{1}
&=&\sigma_{1}\otimes 2\cos{\theta}
=(\sigma_{+}+\sigma_{-})\otimes (e^{i\theta}+e^{-i\theta}) \\
&=&\sigma_{+}\otimes e^{i\theta}+\sigma_{+}\otimes e^{-i\theta}
+\sigma_{-}\otimes e^{i\theta}+\sigma_{-}\otimes e^{-i\theta}
\longrightarrow 
\sigma_{+}\otimes e^{i\theta}+\sigma_{-}\otimes e^{-i\theta}.
\end{eqnarray*}

\vspace{3mm}
In order to solve (\ref{eq:reduced Schrodinger}) we set 
$\hbar=1$ for simplicity. 
From (\ref{eq:reduced Hamiltonian}) it is easy to see
\[
\left(
\begin{array}{cc}
-\frac{\Delta}{2}          & ge^{i(\omega t+\phi)}  \\
ge^{-i(\omega t+\phi)}  &  \frac{\Delta}{2}
\end{array}
\right)
=
\left(
\begin{array}{cc}
e^{i\frac{(\omega t+\phi)}{2}} &       \\
   & e^{-i\frac{(\omega t+\phi)}{2}}
\end{array}
\right)
\left(
\begin{array}{cc}
-\frac{\Delta}{2} & g                      \\
g                      & \frac{\Delta}{2} 
\end{array}
\right)
\left(
\begin{array}{cc}
e^{-i\frac{(\omega t+\phi)}{2}} &     \\
   & e^{i\frac{(\omega t+\phi)}{2}}
\end{array}
\right),
\]
so we transforn the wave function $\Psi$ in 
(\ref{eq:reduced Schrodinger}) into
\begin{equation}
\label{eq:transformation}
\Phi=
\left(
\begin{array}{cc}
e^{-i\frac{(\omega t+\phi)}{2}} &     \\
   & e^{i\frac{(\omega t+\phi)}{2}}
\end{array}
\right)
\Psi 
\ \Longleftrightarrow\ 
\Psi =
\left(
\begin{array}{cc}
e^{i\frac{(\omega t+\phi)}{2}} &       \\
   & e^{-i\frac{(\omega t+\phi)}{2}}
\end{array}
\right)
\Phi. 
\end{equation}
Then the Schr{\"o}dinger equation (\ref{eq:reduced Schrodinger}) 
becomes
\begin{equation}
\label{eq:deformed Schrodinger}
i\frac{\partial}{\partial t}\Phi = 
\left(
\begin{array}{cc}
-\frac{\Delta-\omega}{2} & g                                \\
g                                 & \frac{\Delta-\omega}{2} 
\end{array}
\right)\Phi
\end{equation}
by a straightforward calculation.

\vspace{3mm}
Here we set the resonance condition
\begin{equation}
\label{eq:resonance}
\Delta=\omega\quad (\Longleftrightarrow 
E_{1}-E_{0}=\hbar\omega\quad\mbox{precisely}).
\end{equation}
Namely, we subject the laser field with $\omega$ equal to 
the energy difference $\Delta$. See the following figure.

\vspace{5mm}
\begin{center}
\unitlength 0.1in
\begin{picture}( 21.1000,  5.3500)(  1.1000, -8.2500)
%
\special{pn 8}%
\special{pa 1380 816}%
\special{pa 2220 816}%
\special{fp}%
\put(10.7000,-7.9500){\makebox(0,0){$E_{0}$}}%
\put(10.9000,-3.8000){\makebox(0,0){$E_{1}$}}%
\put(25.0000,-7.8500){\makebox(0,0){$|0\rangle$}}%
\put(24.8000,-3.8000){\makebox(0,0){$|1\rangle$}}%
%
\special{pn 8}%
\special{pa 1360 370}%
\special{pa 2200 370}%
\special{fp}%
%
\special{pn 8}%
\special{pa 612 606}%
\special{pa 588 638}%
\special{pa 566 656}%
\special{pa 546 646}%
\special{pa 528 612}%
\special{pa 512 572}%
\special{pa 496 538}%
\special{pa 482 526}%
\special{pa 468 544}%
\special{pa 454 580}%
\special{pa 440 618}%
\special{pa 424 642}%
\special{pa 406 638}%
\special{pa 386 612}%
\special{pa 366 576}%
\special{pa 344 544}%
\special{pa 326 532}%
\special{pa 306 542}%
\special{pa 290 572}%
\special{pa 272 612}%
\special{pa 260 642}%
\special{sp}%
%
\special{pn 8}%
\special{pa 622 598}%
\special{pa 732 598}%
\special{fp}%
\special{sh 1}%
\special{pa 732 598}%
\special{pa 666 578}%
\special{pa 678 598}%
\special{pa 664 618}%
\special{pa 732 598}%
\special{fp}%
\put(4.7000,-3.7500){\makebox(0,0){$\omega$}}%
%
\special{pn 8}%
\special{pa 1796 596}%
\special{pa 1796 376}%
\special{fp}%
\special{sh 1}%
\special{pa 1796 376}%
\special{pa 1776 442}%
\special{pa 1796 428}%
\special{pa 1816 442}%
\special{pa 1796 376}%
\special{fp}%
%
\special{pn 8}%
\special{pa 1796 606}%
\special{pa 1796 826}%
\special{fp}%
\special{sh 1}%
\special{pa 1796 826}%
\special{pa 1816 758}%
\special{pa 1796 772}%
\special{pa 1776 758}%
\special{pa 1796 826}%
\special{fp}%
\put(32.5000,-6.0000){\makebox(0,0){$E_{1}-E_{0}=\hbar\omega$}}%
\end{picture}%

\end{center}

\vspace{3mm}\noindent
Then (\ref{eq:deformed Schrodinger}) becomes
\[
i\frac{\partial}{\partial t}\Phi 
= 
\left(
\begin{array}{cc}
0 & g \\
g & 0 
\end{array}
\right)\Phi 
=
g\sigma_{1}\Phi
\]
and we have only to solve the equation
\[
\frac{\partial}{\partial t}\Phi=-ig\sigma_{1}\Phi.
\]
By (\ref{eq:exponential}) ($\lambda=-gt$) the solution is
\[
\Phi(t)=e^{-igt\sigma_{1}}\Phi(0)=
\left(
\begin{array}{cc}
\cos(gt)  & -i\sin(gt) \\
-i\sin(gt) & \cos(gt)
\end{array}
\right)
\Phi(0),
\]
and coming back to $\Psi$ (from $\Phi$) we obtain
\begin{eqnarray}
\label{eq:}
\Psi(t)
&=&
\left(
\begin{array}{cc}
e^{i\frac{(\omega t+\phi)}{2}} &       \\
   & e^{-i\frac{(\omega t+\phi)}{2}}
\end{array}
\right)
\left(
\begin{array}{cc}
\cos(gt)  & -i\sin(gt) \\
-i\sin(gt) & \cos(gt)
\end{array}
\right)
\Psi(0) \nonumber \\
&=&
e^{i\frac{(\omega t+\phi)}{2}} 
\left(
\begin{array}{cc}
1 &       \\
   & e^{-i(\omega t+\phi)}
\end{array}
\right)
\left(
\begin{array}{cc}
\cos(gt)  & -i\sin(gt) \\
-i\sin(gt) & \cos(gt)
\end{array}
\right)
\Psi(0) \nonumber \\
&=&
\left(
\begin{array}{cc}
\cos(gt)                              & -i\sin(gt)                             \\
-ie^{-i(\omega t+\phi)}\sin(gt) & e^{-i(\omega t+\phi)}\cos(gt)
\end{array}
\right)
\Psi(0) 
\end{eqnarray}
by (\ref{eq:transformation}) ($\Psi(0)=\Phi(0)$) because 
the total phase $e^{i\frac{(\omega t+\phi)}{2}}$ can be 
neglected in Quantum Mechanics.

As an initial condition, if we choose
\[
\Psi(0)=\ket{0}=
\left(
\begin{array}{@{\,}c@{\,}}
1 \\
0
\end{array}
\right)
\]
we have
\begin{eqnarray}
\Psi(t)
&=&
\left(
\begin{array}{@{\,}c@{\,}}
\cos(gt)                               \\
-ie^{-i(\omega t+\phi)}\sin(gt)
\end{array}
\right)
=
\left(
\begin{array}{@{\,}c@{\,}}
\cos(gt)                                     \\
e^{-i(\omega t+\phi+\pi/2)}\sin(gt)
\end{array}
\right)  \nonumber \\
&=&
\cos(gt)\ket{0}+e^{-i(\omega t+\phi+\pi/2)}\sin(gt)\ket{1}
\end{eqnarray}
by (\ref{eq:basis}). That is, $\Psi(t)$ oscillates between 
the two states $\ket{0}$ and $\ket{1}$. 
This is called the coherent oscillation or the Rabi oscillation, 
which plays an essential role in Quantum Optics.

Concerning an application of this oscillation to Quantum 
Computation see for example \cite{AH}.

\vspace{5mm}\noindent
{\bf Problem}\ \ Our real target is to solve the Schr{\"o}dinger 
equation
\[
i\hbar\frac{\partial}{\partial t}\Psi = H\Psi
\]
with
\[
H=H(t)=
\left(
\begin{array}{cc}
-\frac{\Delta}{2}          & 2g\cos(\omega t+\phi)  \\
2g\cos(\omega t+\phi) & \frac{\Delta}{2} 
\end{array}
\right).
\]
Present a new idea and solve the equation.

\section{Quantum Rabi Model}
In this section we discuss the quantum Rabi model  
whose Hamiltonian is given by
\begin{eqnarray}
\label{Rabi Hamiltonian}
H
&=&\frac{\Omega}{2}\sigma_{3}\otimes {\bf 1}+
\omega{1_{2}}\otimes a^{\dagger}a+
g\sigma_{1}\otimes (a+a^{\dagger})  \nonumber \\
&=&\frac{\Omega}{2}\sigma_{3}\otimes {\bf 1}+
\omega{1_{2}}\otimes N+
g(\sigma_{+}+\sigma_{-})\otimes (a+a^{\dagger}) 
\end{eqnarray}
where ${\bf 1}$ is the identity operator on the Fock space 
${\cal F}$ generated by the Heisenberg algebra 
$\{a, a^{\dagger}, N\equiv a^{\dagger}a\}$, 
and $\Omega$ and $\omega$ are constant, and $g$ is 
a coupling constant. As a general introduction to this 
model see for example see \cite{WS}.

Let us recall the fundamental relations of the Heisenberg algebra
\begin{equation}
\label{eq:fundamental equations}
[N,a^{\dagger}]=a^{\dagger},\quad
[N,a]=-a,\quad
[a,a^{\dagger}]={\bf 1}.
\end{equation}
Here, the Fock space ${\cal F}$ is a Hilbert space over 
$\fukuso$ given by
\[
{\cal F}=\mbox{Vect}_{\fukuso}\left\{\ket{0},\ket{1},
\cdots,\ket{n},\cdots\right\}
\]
where $\ket{0}$ is the vacuum ($a\ket{0}=0$) and 
$\ket{n}$ is given by
\[
\ket{n}=\frac{(a^{\dagger})^{n}}{\sqrt{n!}}\ket{0}
\quad \mbox{for}\quad n\geq 0.
\]
On this space the operators (=infinite dimensional matrices) 
$a^{\dagger}$, $a$ and $N$ are represented as
\begin{eqnarray}
\label{eq:creation-annihilation}
&&a=
\left(
\begin{array}{ccccc}
0 & 1 &          &          &        \\
  & 0 & \sqrt{2} &          &        \\
  &   & 0        & \sqrt{3} &        \\
  &   &          & 0        & \ddots \\
  &   &          &          & \ddots
\end{array}
\right),\quad
a^{\dagger}=
\left(
\begin{array}{ccccc}
0 &          &          &        &        \\
1 & 0        &          &        &        \\
  & \sqrt{2} & 0        &        &        \\
  &          & \sqrt{3} & 0      &        \\
  &          &          & \ddots & \ddots
\end{array}
\right),
\nonumber \\
&&N=a^{\dagger}a=
\left(
\begin{array}{ccccc}
0 &   &   &   &        \\
  & 1 &   &   &        \\
  &   & 2 &   &        \\
  &   &   & 3 &        \\
  &   &   &   & \ddots
\end{array}
\right)
\end{eqnarray}
by use of (\ref{eq:fundamental equations}).

\vspace{5mm}\noindent
{\bf Note}\ \ We can add a phase to $\{a,a^{\dagger}\}$ like
\[
b=e^{i\theta}a,\quad 
b^{\dagger}=e^{-i\theta}a^{\dagger},\quad 
N=b^{\dagger}b=a^{\dagger}a
\]
where $\theta$ is constant. Then we have another Heisenberg algebra
\[
[N,b^{\dagger}]=b^{\dagger},\quad [N,b]=-b,\quad [b,b^{\dagger}]={\bf 1}.
\]

Again, we would like to solve Schr{\"o}dinger equation 
($\hbar=1$ for simplicity)
\begin{equation}
\label{eq:quantum Schrodinger}
i\frac{\partial}{\partial t}\ket{\Psi}
=H\ket{\Psi}
=\left\{\frac{\Omega}{2}\sigma_{3}\otimes {\bf 1}+
\omega{1_{2}}\otimes N+
g(\sigma_{+}+\sigma_{-})\otimes (a+a^{\dagger})\right\}\ket{\Psi}
\end{equation}
exactly. To the best of our knowledge the exact solution 
has not been known, so we must use some approximation 
in order to obatin an analytic solution.

Since
\[
(\sigma_{+}+\sigma_{-})\otimes (a+a^{\dagger})=
\sigma_{+}\otimes a+\sigma_{+}\otimes a^{\dagger}+
\sigma_{-}\otimes a+\sigma_{-}\otimes a^{\dagger},
\]
we neglect the middle terms $\sigma_{+}\otimes a^{\dagger}+
\sigma_{-}\otimes a$ and set
\begin{equation}
\label{eq:Jaynes-Cummings hamiltonian}
\widetilde{H}=\frac{\Omega}{2}\sigma_{3}\otimes {\bf 1}+
\omega{1_{2}}\otimes N+
g(\sigma_{+}\otimes a+\sigma_{-}\otimes a^{\dagger}).
\end{equation}
This is called the rotating wave approximation and 
the resultant Hamiltonian is called 
the Jaynes-Cummings one\footnote{In \cite{WS} 
it is called the Jaynes-Cummings-Paul one}, \cite{JC}. 

Therefore, our modified task is to solve the Schr{\"o}dinger 
equation
\begin{equation}
\label{eq:reduced quantum Schrodinger}
i\frac{\partial}{\partial t}\ket{\Psi}
= \widetilde{H}\ket{\Psi}
=\left\{\frac{\Omega}{2}\sigma_{3}\otimes {\bf 1}+
\omega{1_{2}}\otimes N+
g(\sigma_{+}\otimes a+\sigma_{-}\otimes a^{\dagger})\right\}\ket{\Psi}
\end{equation}
exactly. Mysteriously enough, to solve the equation is very easy.

For a unitary operator $U=U(t)$ we set
\[
\ket{\Phi}=U\ket{\Psi}.
\]
Then it is easy to see
\[
i\frac{\partial}{\partial t}\ket{\Phi}=
\left(U\widetilde{H}U^{-1}+i\frac{\partial U}{\partial t}U^{-1}\right)\ket{\Phi}
\]
by (\ref{eq:reduced quantum Schrodinger}). If we choose $U$ as
\[
U(t)
=e^{it\frac{\omega}{2}\sigma_{3}}\otimes e^{it\omega N}
=
\left(
\begin{array}{cc}
e^{it(\omega N+\frac{\omega}{2})} &   \\
  & e^{it(\omega N-\frac{\omega}{2})}
\end{array}
\right)
\]
(we use $\frac{\omega}{2}$ in place of $\frac{\omega}{2}{\bf 1}$ 
for simplicity), a straightforward calculation gives
\begin{equation}
\label{eq:}
U\widetilde{H}U^{-1}+i\frac{\partial U}{\partial t}U^{-1}
=
\left(
\begin{array}{cc}
\frac{\Omega-\omega}{2} & ga                 \\
ga^{\dagger} & -\frac{\Omega-\omega}{2} 
\end{array}
\right)
\end{equation}
and we have a simple equation
\begin{equation}
\label{eq:quantum reduced}
i\frac{\partial}{\partial t}\ket{\Phi}
=
\left(
\begin{array}{cc}
\frac{\Omega-\omega}{2} & ga                 \\
ga^{\dagger} & -\frac{\Omega-\omega}{2} 
\end{array}
\right)
\ket{\Phi}.
\end{equation}
Note that in the process of calculation we have used 
the relations
\[
e^{it\omega N}a e^{-it\omega N}=e^{-it\omega}a, 
\quad
e^{it\omega N}a^{\dagger}e^{-it\omega N}=e^{it\omega}a^{\dagger}.
\quad
\]
The proof is easy by use of the formula
\begin{equation}
\label{eq:BCH formula}
e^{X}Ae^{-X}=A+[X,A]+\frac{1}{2!}[X,[X,A]]+\frac{1}{3!}[X,[X,[X,A]]]+\cdots
\end{equation}
for square matrices $X$, $A$ and (\ref{eq:fundamental equations}).

\vspace{3mm}
Here we set the resonance condition
\begin{equation}
\label{eq:quantum resonance}
\Omega=\omega,
\end{equation}
then (\ref{eq:quantum reduced}) becomes
\[
i\frac{\partial}{\partial t}\ket{\Phi}
=
\left(
\begin{array}{cc}
 & ga               \\
ga^{\dagger} &  
\end{array}
\right)
\ket{\Phi}
=
g
\left(
\begin{array}{cc}
 & a               \\
a^{\dagger} &  
\end{array}
\right)
\ket{\Phi}.
\]
Let us solve this equation. By setting
\[
A=
\left(
\begin{array}{cc}
                & a  \\
a^{\dagger} &  
\end{array}
\right)
\]
we calculate the term $e^{-igtA}$. Noting
\[
A^{2}=
\left(
\begin{array}{cc}
aa^{\dagger} &                   \\
                  & a^{\dagger}a
\end{array}
\right)=
\left(
\begin{array}{cc}
N+1 &     \\
      & N
\end{array}
\right)
\quad (\Longleftarrow [a,a^{\dagger}]={\bf 1})
\]
we have
\begin{eqnarray}
\label{eq:Q-exponential}
e^{-igt A}
&=&
\sum_{n=0}^{\infty}\frac{(-igt)^{n}A^{n}}{n!}
\nonumber \\
&=&
\sum_{n=0}^{\infty}\frac{(-igt)^{2n}}{(2n)!}A^{2n}+
\sum_{n=0}^{\infty}\frac{(-igt)^{2n+1}}{(2n+1)!}A^{2n+1}
\nonumber \\
&=&
\sum_{n=0}^{\infty}\frac{(-1)^{n}(gt)^{2n}}{(2n)!}
\left(
\begin{array}{cc}
(N+1)^{n} &         \\
             & N^{n}
\end{array}
\right)
-i
\sum_{n=0}^{\infty}\frac{(-1)^{n}(gt)^{2n+1}}{(2n+1)!}
\left(
\begin{array}{cc}
                       & (N+1)^{n}a  \\
N^{n}a^{\dagger} &  
\end{array}
\right)   \nonumber \\
&=&
\left(
\begin{array}{cc}
\cos(\sqrt{N+1}gt) &                       \\
                        & \cos(\sqrt{N}gt)
\end{array}
\right)
-i
\left(
\begin{array}{cc}
   & \frac{1}{\sqrt{N+1}}\sin(\sqrt{N+1}gt)a      \\
\frac{1}{\sqrt{N}}\sin(\sqrt{N}gt)a^{\dagger} & 
\end{array}
\right)   \nonumber \\
&=&
\left(
\begin{array}{cc}
\cos(\sqrt{N+1}gt) & -i\frac{1}{\sqrt{N+1}}\sin(\sqrt{N+1}gt)a      \\
-i\frac{1}{\sqrt{N}}\sin(\sqrt{N}gt)a^{\dagger} & \cos(\sqrt{N}gt)
\end{array}
\right).
\end{eqnarray}
Therefore, the solution is given by
\[
\ket{\Phi(t)}=
\left(
\begin{array}{cc}
\cos(\sqrt{N+1}gt) & -i\frac{1}{\sqrt{N+1}}\sin(\sqrt{N+1}gt)a      \\
-i\frac{1}{\sqrt{N}}\sin(\sqrt{N}gt)a^{\dagger} & \cos(\sqrt{N}gt)
\end{array}
\right)
\ket{\Phi(0)}
\]
and coming back to $\ket{\Psi}$ (from $\ket{\Phi}$) we finally obtain
\begin{eqnarray}
\ket{\Psi(t)}=
\left(
\begin{array}{cc}
e^{it(\omega N+\frac{\omega}{2})} &   \\
  & e^{it(\omega N-\frac{\omega}{2})}
\end{array}
\right)
\left(
\begin{array}{cc}
\cos(\sqrt{N+1}gt) & -i\frac{1}{\sqrt{N+1}}\sin(\sqrt{N+1}gt)a      \\
-i\frac{1}{\sqrt{N}}\sin(\sqrt{N}gt)a^{\dagger} & \cos(\sqrt{N}gt)
\end{array}
\right)
\ket{\Psi(0)}  \nonumber \\
&{}&
\end{eqnarray}
where $\ket{\Psi(0)}=\ket{\Phi(0)}$.

As an initial condition, if we choose
\[
\ket{\Psi(0)}=
\left(
\begin{array}{@{\,}c@{\,}}
1 \\
0
\end{array}
\right)\otimes \ket{0}
=
\left(
\begin{array}{@{\,}c@{\,}}
\ket{0} \\
0
\end{array}
\right)
\]
we have
\begin{equation}
\ket{\Psi(t)}=
\left(
\begin{array}{@{\,}c@{\,}}
e^{it\frac{\omega}{2}}\cos(gt)\ket{0}   \\
-ie^{it\frac{\omega}{2}}\sin(gt)\ket{1}
\end{array}
\right)
\end{equation}
because $a\ket{0}=0$ and $a^{\dagger}\ket{0}=\ket{1}$,  
or
\[
\ket{\Psi(t)}=
\cos(gt)
\left(
\begin{array}{@{\,}c@{\,}}
1 \\
0
\end{array}
\right)\otimes \ket{0}
+
e^{-i\frac{\pi}{2}}\sin(gt)
\left(
\begin{array}{@{\,}c@{\,}}
0 \\
1
\end{array}
\right)\otimes \ket{1}
\]
where the total phase $e^{it\frac{\omega}{2}}$ has been removed.

\vspace{3mm}\noindent
{\bf Problem}\ \ Our real target is to solve the Schr{\"o}dinger 
equation
\[
i\hbar\frac{\partial}{\partial t}\ket{\Psi} = H\ket{\Psi}
\]
with
\[
H=H(t)=
\frac{\Omega}{2}\sigma_{3}\otimes {\bf 1}+\omega{1_{2}}\otimes N+
g(\sigma_{+}+\sigma_{-})\otimes (a+a^{\dagger}) 
\]
Present a new idea and solve the equation.

\vspace{3mm}
As a developed version of the Jaynes-Cummings model 
see for example \cite{FS-1} and \cite{FS-2}.

\section{Concluding Remarks}
In this note I introduced the rotating wave approximation 
which plays an important role in Quantum Optics with 
two examples. 
The problem is that the method is used even in a subtle 
case. As far as I know it is very hard to obtain an 
analytic approximate solution without RWA.

I don't know the reason why it is so. However, such a 
``temporary" method must be overcome in the near future. 
I expect that young researchers will attack and overcome 
this problem.

Concerning a recent criticism to RWA see \cite{JL} and 
its references, and concerning recent applications 
to the dynamical Casimir effect see \cite{Law}, \cite{Dods} 
and \cite{FS-3}, \cite{FS-4} (\cite{FS-3} and \cite{FS-4} 
are highly recommended).

\vspace{10mm}
\begin{center}
\begin{Large}
{\bf Appendix}
\end{Large}
\end{center}

\vspace{10mm}\noindent
{\bf [A]\ \ Another Approach}

\noindent
Let us give another approach to the derivation 
(\ref{eq:exponential}), which may be smart enough. 
It is easy to see the diagonal form
\[
\sigma_{1}=W\sigma_{3}W^{-1}
\]
where $W$ is the Walsh--Hadamard matrix (operation) 
given by
\[
W=\frac{1}{\sqrt{2}}
\left(
\begin{array}{cc}
1 & 1  \\
1 & -1
\end{array}
\right)\ \in \ O(2).
\]
Note that
\[
W^{2}=1_{2} \Longrightarrow W=W^{-1}.
\]
Then we obtain
\begin{eqnarray*}
e^{i\lambda \sigma_{1}}
&=&
e^{i\lambda W\sigma_{3}W^{-1}}=We^{i\lambda \sigma_{3}}W^{-1} \\
&=&
\frac{1}{2}
\left(
\begin{array}{cc}
1 & 1  \\
1 & -1
\end{array}
\right)
\left(
\begin{array}{cc}
e^{i\lambda} &                     \\
                 & e^{-i\lambda}
\end{array}
\right)
\left(
\begin{array}{cc}
1 & 1  \\
1 & -1
\end{array}
\right)      \\
&=&
\left(
\begin{array}{cc}
\frac{e^{i\lambda}+e^{-i\lambda}}{2} & 
\frac{e^{i\lambda}-e^{-i\lambda}}{2}       \\
\frac{e^{i\lambda}-e^{-i\lambda}}{2} & 
\frac{e^{i\lambda}+e^{-i\lambda}}{2} 
\end{array}
\right) \\
&=&
\left(
\begin{array}{cc}
\cos{\lambda} & i\sin{\lambda}  \\
i\sin{\lambda} & \cos{\lambda}
\end{array}
\right).
\end{eqnarray*}

Readers should remark that the Walsh--Hadamard 
matrix $W$ plays an essential role in Quantum Computation. 
See for example \cite{AB} (note : $W\rightarrow U_{A}$ in 
this paper).

\vspace{10mm}\noindent
{\bf [B]\ \ Tensor Product}

\noindent
Let us give a brief introduction to the tensor product 
of matrices. 
For $A=(a_{ij})\in M(m;\fukuso)$ and $B=(b_{ij})\in M(n;\fukuso)$ 
the tensor product is defined by
\[
A\otimes B=(a_{ij})\otimes B=(a_{ij}B)\in M(mn;\fukuso).
\]
Precisely, in case of $m=2$ and $n=3$
\begin{eqnarray*}
A\otimes B
&=&
\left(
\begin{array}{cc}
a_{11} & a_{12} \\
a_{21} & a_{22}
\end{array}
\right)\otimes B
=
\left(
\begin{array}{cc}
a_{11}B & a_{12}B \\
a_{21}B & a_{22}B
\end{array}
\right)  \\
&=&
\left(
\begin{array}{cccccc}
a_{11}b_{11} & a_{11}b_{12} & a_{11}b_{13} & a_{12}b_{11} & a_{12}b_{12} & a_{12}b_{13} \\
a_{11}b_{21} & a_{11}b_{22} & a_{11}b_{23} & a_{12}b_{21} & a_{12}b_{22} & a_{12}b_{23} \\
a_{11}b_{31} & a_{11}b_{32} & a_{11}b_{33} & a_{12}b_{31} & a_{12}b_{32} & a_{12}b_{33} \\
a_{21}b_{11} & a_{21}b_{12} & a_{21}b_{13} & a_{22}b_{11} & a_{22}b_{12} & a_{22}b_{13} \\
a_{21}b_{21} & a_{21}b_{22} & a_{21}b_{23} & a_{22}b_{21} & a_{22}b_{22} & a_{22}b_{23} \\
a_{21}b_{31} & a_{21}b_{32} & a_{21}b_{33} & a_{22}b_{31} & a_{22}b_{32} & a_{22}b_{33}
\end{array}
\right).
\end{eqnarray*}

When I was a young student in Japan this product 
was called the Kronecker one. Nowadays, it is called 
the tensor product in a unified manner, which may 
be better.

Note that
\[
1_{2}\otimes B
=
\left(
\begin{array}{cccccc}
b_{11} & b_{12} & b_{13} &   &   &   \\
b_{21} & b_{22} & b_{23} &   &   &   \\
b_{31} & b_{32} & b_{33} &   &   &   \\
  &   &   & b_{11} & b_{12} & b_{13} \\
  &   &   & b_{21} & b_{22} & b_{23} \\
  &   &   & b_{31} & b_{32} & b_{33}
\end{array}
\right),
\]
while
\[
B\otimes 1_{2}
=
\left(
\begin{array}{cccccc}
b_{11} &   & b_{12} &   & b_{13} &   \\
 & b_{11} &   & b_{12} &   & b_{13}  \\
b_{21} &   & b_{22} &   & b_{23} &   \\
 & b_{21} &   & b_{22} &   & b_{23}  \\
b_{31} &   & b_{32} &   & b_{33} &   \\
 & b_{31} &   & b_{32} &   & b_{33}
\end{array}
\right).
\]
The blanks in the matrices above are of course zero. 

Readers should recognize the difference. See for example \cite{Five} 
for more details.

\vspace{10mm}\noindent
{\bf [C]\ \ Beyond the RWA}

\noindent
Let us try to solve the equation (\ref{eq:Schrodinger}). 
For the purpose it is convenient to assume a form
for some solution
\[
\Psi(t)=e^{-iF(t)\sigma_{+}}e^{-iG(t)\tau_{3}}e^{-iH(t)\sigma_{-}}\Psi(0),
\quad F(0)=G(0)=H(0)=0
\]
where we set $\tau_{3}=(1/2)\sigma_{3}$ for simplicity.
Note that this form called the disentangling form (a kind of 
Gauss decomposition of some matrices) is very popular 
in Quantum Physics.

By setting $\hbar=1$ for simplicity in (\ref{eq:Schrodinger}) 
we must calculate 
\begin{eqnarray*}
i\frac{\partial}{\partial t}\Psi 
&=& 
\left\{
-\Delta\tau_{3}+2g\cos(\omega t+\phi)(\sigma_{+}+\sigma_{-})
\right\}\Psi  \\
&=&
\left\{2g\cos(\omega t+\phi)\sigma_{+}-\Delta\tau_{3}+
2g\cos(\omega t+\phi)\sigma_{-}
\right\}\Psi
\end{eqnarray*}
where $\tau_{3}=(1/2)\sigma_{3}$. 

Then we have
\begin{eqnarray*}
i\frac{\partial}{\partial t}\Psi
&=& 
i\frac{\partial}{\partial t}
\left\{e^{-iF(t)\sigma_{+}}e^{-iG(t)\tau_{3}}e^{-iH(t)\sigma_{-}}\right\}\Psi(0) \\
&=&
\left\{
\dot{F}(t)\sigma_{+}+\dot{G}(t)e^{-iF(t)\sigma_{+}}\tau_{3}e^{iF(t)\sigma_{+}}+
\dot{H}(t)e^{-iF(t)\sigma_{+}}e^{-iG(t)\tau_{3}}\sigma_{-}e^{iG(t)\tau_{3}}e^{iF(t)\sigma_{+}}
\right\}\Psi.
\end{eqnarray*}
From (\ref{eq:su(2)-relations})
\[
[\tau_{3},\sigma_{+}]=\sigma_{+},\quad
[\tau_{3},\sigma_{-}]=-\sigma_{-},\quad
[\sigma_{+},\sigma_{-}]=2\tau_{3}
\]
and the formula (\ref{eq:BCH formula}) it is easy to see
\begin{eqnarray*}
&&e^{-iF(t)\sigma_{+}}\tau_{3}e^{iF(t)\sigma_{+}}=\tau_{3}+iF(t)\sigma_{+}, \\
&&e^{-iG(t)\tau_{3}}\sigma_{-}e^{iG(t)\tau_{3}}=(1+iG(t))\sigma_{-},\quad 
e^{-iF(t)\sigma_{+}}\sigma_{-}e^{iF(t)\sigma_{+}}=\sigma_{-}-2iF(t)\tau_{3}+F(t)^{2}\sigma_{+}.
\end{eqnarray*}
Therefore
\begin{eqnarray*}
i\frac{\partial}{\partial t}\Psi
&=& 
\left\{
\dot{F}(t)\sigma_{+}+
\dot{G}(t)(\tau_{3}+iF(t)\sigma_{+})+
\dot{H}(t)(1+iG(t))(\sigma_{-}-2iF(t)\tau_{3}+F(t)^{2}\sigma_{+})
\right\}\Psi \\
&=&
\left[
\left\{\dot{F}(t)+iF(t)\dot{G}(t)+(1+iG(t))F(t)^{2}\dot{H}(t)\right\}\sigma_{+}+
\right. \\
&&
\left. \
\left\{\dot{G}(t)-2i(1+iG(t))F(t)\dot{H}(t)\right\}\tau_{3}+
(1+iG(t))\dot{H}(t)\sigma_{-}
\right]\Psi.
\end{eqnarray*}

By comparing two equations above we obtain a system of 
differential equations
\[
\left\{
\begin{array}{lll}
\dot{F}(t)+iF(t)\dot{G}(t)+(1+iG(t))F(t)^{2}\dot{H}(t)=2g\cos(\omega t+\phi), \\
\dot{G}(t)-2i(1+iG(t))F(t)\dot{H}(t)=-\Delta, \\
(1+iG(t))\dot{H}(t)=2g\cos(\omega t+\phi).
\end{array}
\right.
\]
By deforming them we have
\[
\left\{
\begin{array}{lll}
\dot{F}(t)-i\Delta F(t)-2g\cos(\omega t+\phi)F(t)^{2}=2g\cos(\omega t+\phi), \\
\dot{G}(t)-4ig\cos(\omega t+\phi)F(t)=-\Delta, \\
(1+iG(t))\dot{H}(t)=2g\cos(\omega t+\phi).
\end{array}
\right.
\]
This is a simple exercise for young students. 

If we can solve the first equation then we obtain solutions like
\[
F(t)\Longrightarrow G(t)\Longrightarrow H(t).
\]
The first equation
\[
\dot{F}(t)-2g\cos(\omega t+\phi)-i\Delta F(t)-2g\cos(\omega t+\phi)F(t)^{2}=0
\]
is a (famous) {\bf Riccati equation} of general type. Unfortunately, 
we don't know how to solve it explicitly at the present time.

\vspace{10mm}\noindent
{\bf [D]\ \ Full Calculation}

\noindent
Let us give the full calculation to the equation 
(\ref{eq:quantum reduced}). We set
\[
B=
\left(
\begin{array}{cc}
\frac{\Omega-\omega}{2} & ga                 \\
ga^{\dagger} & -\frac{\Omega-\omega}{2} 
\end{array}
\right)
\]
and calculate $e^{-itB}$ without assuming $\Omega=\omega$ 
in (\ref{eq:quantum resonance}). 
Again, noting
\[
B^{2}
=
\left(
\begin{array}{cc}
\left(\frac{\Omega-\omega}{2}\right)^{2}+g^{2}aa^{\dagger}  &   \\
  & \left(\frac{\Omega-\omega}{2}\right)^{2}+g^{2}a^{\dagger}a
\end{array}
\right)
=
\left(
\begin{array}{cc}
\left(\frac{\Omega-\omega}{2}\right)^{2}+g^{2}N+g^{2} &   \\
  & \left(\frac{\Omega-\omega}{2}\right)^{2}+g^{2}N
\end{array}
\right)
\]
($aa^{\dagger}=a^{\dagger}a +1=N+1$) we obtain
\begin{eqnarray*}
e^{-itB}
&=&
\exp
\left\{-it
\left(
\begin{array}{cc}
\frac{\Omega-\omega}{2} & ga               \\
ga^{\dagger} & -\frac{\Omega-\omega}{2}
\end{array}
\right)
\right\} \nonumber \\
&=&
\left(
\begin{array}{cc}
\cos t\sqrt{\varphi+g^{2}}-\frac{i\delta}{2}\frac{\sin t\sqrt{\varphi+g^{2}}}{\sqrt{\varphi+g^{2}}} & 
-ig\frac{\sin t\sqrt{\varphi+g^{2}}}{\sqrt{\varphi+g^{2}}}a \\
-ig\frac{\sin t\sqrt{\varphi}}{\sqrt{\varphi}}a^{\dagger} &
\cos t\sqrt{\varphi}+\frac{i\delta}{2}\frac{\sin t\sqrt{\varphi}}{\sqrt{\varphi}}
\end{array}
\right)
\end{eqnarray*}
where we have set
\[
\delta\equiv \Omega-\omega,\quad 
\varphi\equiv \frac{\delta^{2}}{4}+g^{2}N
\]
for simplicity. See (\ref{eq:Q-exponential}). 
This is a good exercise for young students.


%

\end{document}